\documentclass[aps,twocolumn,10pt,aps,amsmath,amssymb,nofootinbib,superscriptaddress,showkeys,showpacs]{revtex4-1}
\usepackage{epsfig} 
\usepackage{amsmath} 
\usepackage{graphicx}
\usepackage{color}
\usepackage{float}
\usepackage{soul,xcolor}
\usepackage[font=scriptsize]{caption}
\usepackage{braket}
\usepackage{bbold}
\usepackage{subfig}
\usepackage{braket}
\usepackage{titlesec}
\usepackage{placeins}
\usepackage{caption}
\usepackage{bbold}
\usepackage[normalem]{ulem}
\begin{document}
	\title{Effects of nonstandard interaction on temporal and spatial
		correlations in neutrino oscillations}
	
	\author{Trisha Sarkar}
	\email{sarkar.2@iitj.ac.in}
	\affiliation{Indian Institute of Technology Jodhpur, Jodhpur 342037, India}
	
	\author{Khushboo Dixit}
	\email{dixit.1@iitj.ac.in}
	\affiliation{Indian Institute of Technology Jodhpur, Jodhpur 342037, India}

	\date{\today} 
	
	\begin{abstract}
	Effects of physics beyond the standard model in the neutrino sector are conveniently incorporated through non-standard interaction parameters. Assuming new physics in the form of dimension-6 vector operators, a recent global analysis of neutrino oscillation data including results from COHERENT experiment suggests two favorable new physics scenarios. These are LMA-Light (with normal mass ordering) \& LMA-Dark (with inverted mass ordering) sectors of parameters. In this work, we study the effects of new physics solutions on Leggett-Garg-type (LGtI) inequality which quantifies temporal correlations in the system along with flavor entropy and genuine tripartite entanglement which can be considered as measures of spatial correlations. We show that the violation of LGtI for $\nu_{\mu}$ energy around 3 GeV in the DUNE experimental set-up can not only be an indication of presence of new physics but such a new physics is expected to be in the form of LMA-Dark sector with inverted ordering. Further, we show that the LMA-Light solution, in general, decreases the values of all measures of quantum correlations in comparison to their SM predictions. On the other hand, the Dark solution can significantly enhance the values of these measures. 
	
	\end{abstract}
	\maketitle

\section{Introduction}

The currently running experiments at the LHC along with the experiments such as BaBar and Belle have  provided several engrossing evidences of physics beyond the Standard Model (SM) of electroweak interactions. These include hints of Lepton Flavour Universality (LFU) violation in the decays induced by the charged current quark level transition $b \to c l \nu$ ($l=e,\mu,\tau$) \cite{chargedSL} as well as in the neutral current  $b \rightarrow s\,l^+l^-$ ($l=e,\mu$) \cite{Aaij:2017vbb,Aaij:2019wad} decays. The preferred Lorentz structure(s) of the possible new physics \cite{Alguero:2019ptt,Alok:2019ufo,Ciuchini:2019usw,DAmico:2017mtc,Freytsis:2015qca,Alok:2017qsi,Alok:2018uft,Alok:2019uqc,Blanke:2019qrx} can be realized through several extensions of the SM.

The effects of new physics can also manifest in the neutrino sector. The experimental facilities in neutrino physics are now tending towards higher precision and have potential to probe such sub-leading effects. This has triggered a considerable interest in the neutrino physics community. The new physics effects in  neutrino interactions are conveniently incorporated through effective Non-Standard Interaction (NSI) parameters \cite{Grossman:1995wx,Krastev,Brooijmans,Gonzalez,Bergmann,Guzzo2,Guzzo3,Antusch:2008tz,Ohlsson:2012kf,Farzan:2017xzy,Falkowski:2018dmy,Esteban,Esteban3}.

SM can be assumed to be the low energy renormalizable approximation, containing only dimension $D\leq4$ operators, of a complete theory existing at much higher mass scale $\sim\Lambda$, the new physics effects of which can be included in terms of operators having higher dimensional Lorentz structures $(D>4)$ constructed out of SM fermion fields. In this work, we restrict ourselves to dimension-6 vector operators which may show sub-leading effects in long baseline (LBL) neutrino experiments such as Deep Underground Neutrino Experiment (DUNE). 

In a recent analysis, bounds on NSI parameters were obtained by performing a global fit at all relevant data in the neutrino sector. This includes coherent neutrino-nucleus scattering data from COHERENT experiment \cite{Esteban3}. In this analysis, two new physics scenarios have been identified as the most favourable solutions to the global data: 
\begin{enumerate}
	\item LMA-Light sector ($0 < \theta_{12} < \pi/4$) with normal ordering (NO),
	\item LMA-Dark sector ($\pi/4 < \theta_{12} < \pi/2$) with inverted ordering (IO).
\end{enumerate}

These new physics effects can also affect the temporal and spatial correlations present in the system. The most popular criteria to test spatial quantum correlations is Bell's inequality. However, till date, it is not clear how such measurements can be performed in the neutrino sector using the current experimental set-ups. These spatial correlations can also be quantified in terms of flavor entropy \cite{Blasone:2007vw,Dixit:2018kev} and genuine tripartite entanglement \cite{Dixit:2018kev}. These are basically measures of entanglement embedded in the system.

On the other hand, the determination of temporal correlations based on the assumptions of {\it macrorealism} (MR) and {\it noninvasive measurement} (NIM) and usually quantified in terms of {\it Leggett-Garg inequalities} (LGI) is experimentally feasible in the context of neutrino oscillations. In fact, violations of a class of such inequalities, Leggett-Garg-type inequalities (LGtI), using data from MINOS and Daya Bay experiments have been demonstrated in refs. \cite{Formaggio:2016cuh} and \cite{Fu:2017hky}, respectively. The LGtI is constructed by replacing the NIM condition by a weaker condition called stationarity \cite{Huelga}. Such inequalities are more suited for the study of temporal correlations in the neutrino sector in comparison to the LGIs as measurement of neutrinos destroys the NIM assumption. Further, LGtIs can be expressed in terms of neutrino survival and transition probabilities \cite{Formaggio:2016cuh,Naikoo:2017fos}.

In this work we study new physics effects, in particular the impact of two new physics solutions obtained in \cite{Esteban3}, on temporal correlations in neutrino oscillations quantified in terms of LGtI. We intended to identify parameter space where violation of LGtI can provide unambiguous signatures of new physics. Further, we also study NSI effects on flavor entropy and genuine tripartite entanglement present in the neutrino system. Moreover, we also analyze correlations of these observables with the neutrino transition probability. We present our results in the context of upcoming LBL DUNE experimental set-up. We show that the violation of LGtI for $\nu_{\mu}$ energy $\approx$ 3 GeV in the DUNE experimental set-up can not only be an indication of presence of new physics but such a new physics is expected to be in the form of LMA-Dark sector of $\theta_{12}$ with IO.

The new physics effects in the context of quantum correlations were first incorporated in \cite{Dixit:2019cvm} where the NSI effect on a measure of quantum coherence was studied. While this work was in preparation, the article \cite{Shafaq:2020sqo} appeared on the arXiv where NSI effects on LGI was studied. It was shown in \cite{Shafaq:2020sqo} that LGI violation can be enhanced as compared to the standard scenario for specific choices of NSI parameters. In this work we study LGtI under the effects of NSI, however, apart from the study of suppression and enhancement in the value of LGtI parameter over the SM value, we focus on identifying the parameter space where one can get unequivocal imprints of new physics. Additionally, we study NSI effects on spatial correlations as well.

The Plan of this work is as follows. In Sec. \ref{Formalism}, we illustrate the dynamics of neutrino oscillations within SM interaction as well in the presence of NSI. We also define the measures of temporal and spatial quantum correlation used in this work. Then in Sec. \ref{Results}, we present and explain our results. Finally, we conclude in Sec. \ref{Conclusions}.

\section{Formalism} \label{Formalism}

In this section, we present the theoretical framework of our analysis. We start with the dynamics of neutrino oscillations under the effect of both SM interaction and NSI in subsection \ref{SMI} and \ref{NSI}, respectively. Then in subsection \ref{QC} we define the correlation measures used in this work.

\subsection{Neutrino oscillation in matter}\label{SMI}
Let us consider that the neutrino is produced initially in the flavour state $\ket{\nu_{\alpha}}$  $(\alpha=e,\mu,\tau)$ at time $t=0$ . The flavour state is related to the mass eigenstate $\ket{\nu_{i}}$  $(i=1,2,3)$ by the so called 3$\times$3 unitary mixing matrix $(U)$ (PMNS matrix)  as,
\begin{equation}\label{eqn1}
\ket{\nu_{\alpha}}=\sum\limits_{i=1}^3 U_{\alpha i}^{*} \ket{\nu_{i}}.
\end{equation}
Time evolved mass eigenstates at time $t$ can be represented by $\ket{\nu_{i}(t)} = e^{-i\mathcal{H}_m t} \ket{\nu_i} = e^{-i E_i t} \ket{\nu_i}$, where $\mathcal{H}_m$ is the Hamiltonian of neutrino propagation in mass basis and $E_i$ are the eigenvalues corresponding to $\ket{\nu_i}$. Then the time evolution of the flavor state is given as,
\begin{equation}\label{eqn2}
\ket{\nu_{\alpha}(t)}=e^{-i\mathcal{H}_{f}t}\ket{\nu_{\alpha}}=U_{f}(t)\ket{\nu_{\alpha}},
\end{equation}
where $\mathcal{H}_{f} = U\mathcal{H}_{m}U^{\dagger}$ is the Hamiltonian of neutrino oscillation in flavour basis.

The Hamiltonian $H_{f}$ in the flavour basis, when neutrino propagates in matter, is given as
\small
\begin{equation}\label{eqn3}
\mathcal{H}_{f}=\mathcal{H}_{vac}+\mathcal{H}_{mat}=U\begin{pmatrix}
E_{1} & 0 & 0\\
0 & E_{2} & 0\\
0 & 0 & E_{3}
\end{pmatrix}U^{\dagger} + A
\begin{pmatrix}
1 & 0 & 0\\
0 & 0 & 0\\
0 & 0 & 0
\end{pmatrix},
\end{equation}
\normalsize
where $A=\pm\sqrt{2}G_{F}N_{e}$ is standard matter potential, $G_{F}$ is the Fermi constant and $N_{e}$ is the electron number density. The sign of $A$ is positive for neutrinos and negative for anti-neutrinos. Following the framework of \cite{Ohlsson:1999xb}, in the ultra-relativistic limit $t \equiv L$, the flavour evolution operator can be obtained as
\small
\begin{eqnarray}\label{eq}
U_{f}(L) = e^{-i\mathcal{H}_{f}L}
&=&\phi \sum\limits_{a=1}^3 e^{-iL\lambda_{a}} \frac{1}{3\lambda_{a}^{2}+C_{1}}\nonumber\\
&&\times
\left[(\lambda_{a}^{2}+C_{1})I+\lambda_{a}\tilde{T}+\tilde{T}^{2}\right],
\end{eqnarray}
\normalsize
where $T = \mathcal{H}_{m}-tr(\mathcal{H}_{m})I/3$ is the traceless matrix, $\phi=\exp(-iL\,tr(\mathcal{H}_{m})I/3)$ and $\tilde{T}=U\,T\,U^{\dagger}$. Further, $\lambda_{a}$ $(a=1,2,3)$ are the eigenvalues of $T$-matrix and $C_{1}=Det(T)\,tr(T^{-1})$. 

\subsection{Non Standard Interaction in neutrino oscillation}\label{NSI}
In addition to the standard interactions, the neutrino dynamics can also be affected by NSI. Effects of NSI can be more visible for long baseline experiments, such as DUNE, which has the baseline $L \approx 1300$ km and energy range of neutrinos $E = 1 - 10$ GeV \cite{Acciarri:2016crz} (with maximum neutrino-flux in the range $E \approx 3 - 4$ GeV \cite{Abi:2020qib}).
NSI can be classified in two types: charged current (CC)-NSI and neutral current (NC)-NSI. CC-NSI mainly affects neutrino production and detection processes \cite{FernandezMartinez:2011zz,Biggio:2009nt}, while NC-NSI affects the neutrino propagation in matter via coherent forward elastic scattering \cite{Esteban3}. The effect of incoherent scattering is neglected in case of Earth matter density $\rho\sim2.8$ $\rm gm/cc$, as the mean free path for the process is much larger than Earth's diameter when the neutrino energy is lower than $\sim10^5$ GeV \cite{Giunti:2007ry}. The CC-NSI is strictly constrained, at least by an order of magnitude in comparison to the NC-NSI \cite{Biggio:2009nt}, due to bounds coming mainly from the Fermi constant, CKM unitarity, pion decay and the kinematic measurements of the masses of the gauge  bosons $M_Z$ and $M_W$.

SM can be considered the lower energy effective theory of some higher dimensional theory valid at much higher energy scale. Therefore, the effective Lagrangian can be expressed in terms of higher dimensional $(d)$ non-renormalizable operators $(\mathcal{O}_{i,d})$, 
\begin{equation}\label{eqn4}
\mathcal{L}_{eff}=\mathcal{L}_{SM}+\frac{1}{\Lambda}\sum\limits_{i}C_{i,5}\mathcal{O}_{i,5}+\frac{1}{\Lambda^{2}}\sum\limits_{i}C_{i,6}\mathcal{O}_{i,6}+...
\end{equation}
Here $\Lambda$ is the scale of new physics and $C_{i}$'s are the coefficients encapsulating the short-distance physics. Beyond SM, dimension-5 Weinberg operator is the first higher dimensional operator which can generate small neutrino mass after electroweak symmetry breaking. However, the required new physics scale is $\sim 10^{13}$ GeV for the generation of neutrino mass of the order of 1 eV, which is beyond the energy range of LHC \cite{Buchmuller:1985jz,Krauss:2013lra,Babu:2009aq}. Operators of dimension-6 and 8 are studied extensively in \cite{Gavela:2008ra,Antusch:2008tz}. In our work, we are focusing on lepton number conserving dimension-6 four-fermion operators which can significantly affect neutrino oscillations through NSI \cite{Meloni:2009cg}. Lagrangian for CC and NC-NSI are represented using the dimension-6 operators as following \cite{Babu:2019iml,Davidson:2003ha}
\begin{widetext}
	\begin{equation}\label{eq4}
	\begin{split}
	\mathcal{L}_{CC-NSI}=2\sqrt{2}G_{F}\epsilon_{\alpha \beta}^{ff',L}(\Bar{\nu}_{\alpha}\gamma^{\mu}P_{L}l_{\beta})(\Bar{f^{\prime}}\gamma_{\mu}P_{L}f)
	+2\sqrt{2}G_{F}\epsilon_{\alpha \beta}^{ff',R}(\Bar{\nu}_{\alpha}\gamma^{\mu}P_{L}l_{\beta})(\Bar{f^{\prime}}\gamma_{\mu}P_{R}f),
	\\
	\mathcal{L}_{NC-NSI}=2\sqrt{2}G_{F}\epsilon_{\alpha \beta}^{f,L}(\Bar{\nu}_{\alpha}\gamma^{\mu}P_{L}\nu_{\beta})(\Bar{f}\gamma_{\mu}P_{L}f)
	+2\sqrt{2}G_{F}\epsilon_{\alpha \beta}^{f,R}(\Bar{\nu}_{\alpha}\gamma^{\mu}P_{L}\nu_{\beta})(\Bar{f}\gamma_{\mu}P_{R}f).
	\end{split}
	\end{equation}
\end{widetext}
Here, $P_{L,R}=(1\mp \gamma^{5})/2$ are left and right handed chirality operators.  $\epsilon_{\alpha \beta}^{ff'}$ and $\epsilon_{\alpha \beta}^{f}$  are the dimensionless coefficients which give relative strength of NSI for CC and NC, respectively. For CC-NSI, $f\neq f'$ and $f,f'=u,d$ while for NC-NSI, $f=e,u,d$.

The concept of NSI was first introduced in \cite{Wolfenstein:1977ue} in terms of flavour changing neutral current (FCNC) as shown in Eq. (\ref{eq4}). In the limit $\epsilon_{\alpha \beta}^{f}\rightarrow0$, SM result is restored. When $\epsilon_{\alpha \beta}^{f}\sim1$, the new physics effects have the same strength as SM weak interaction. $\epsilon_{\alpha \beta}^{f} \neq 0$ for $\alpha \neq \beta$ implies lepton flavour violation (LFV) and $\epsilon_{\alpha \alpha}^{f} \neq \epsilon_{\beta \beta}^{f}$ shows lepton flavour universality violation (LFUV). For neutrino oscillation in matter, vector part of NSI, $\epsilon_{\alpha\beta}^{f}=\epsilon_{\alpha\beta}^{f,L}+\epsilon_{\alpha\beta}^{f,R}$, is relevant. For detailed review on NSI, see \cite{Dev:2019anc}.

In the presence of NSI, the Hamiltonian in flavour basis given in Eq. (\ref{eqn3}) is modified as,
\small
\begin{equation}\label{eqn5}
\mathcal{H}_{f}=U\begin{pmatrix}
0 & 0 & 0\\
0 & \frac{\Delta m_{21}^{2}}{2E} & 0\\
0 & 0 & \frac{\Delta m_{31}^{2}}{2E}\\
\end{pmatrix} U^{\dagger}+ A \begin{pmatrix}
1+\epsilon_{ee}(x) & \epsilon_{e\mu}(x) & \epsilon_{e\tau}(x)\\
\epsilon_{\mu e}(x) & \epsilon_{\mu \mu}(x) & \epsilon_{\mu \tau}(x)\\
\epsilon_{\tau e}(x) & \epsilon_{\tau \mu}(x) & \epsilon_{\tau \tau}(x)
\end{pmatrix}.
\end{equation}
\normalsize
Here $E$ = $E_{1} + E_{2} + E_{3}$. From Hermiticity condition, $\epsilon_{\alpha \beta}=\epsilon_{\beta \alpha}^{*}$. The off-diagonal terms are in general considered to be complex and can be given as
\begin{equation}\label{eqn6}
\epsilon_{\alpha\beta}=|\epsilon_{\alpha\beta}| e^{i\phi_{\alpha\beta}}.
\end{equation}
The NSI parameters appeared in Eq. (\ref{eqn6}) are related to those in Eq. (\ref{eq4}) as
\begin{equation}\label{eqn7}
\epsilon_{\alpha\beta}=\sum\limits_{f=e,u,d} \frac{N_{f}(x)}{N_{e}(x)} \epsilon_{\alpha\beta}^{f}.
\end{equation}
Here $N_{f}(x)$ is the fermion density and $x$ is the distance travelled by neutrino in matter. From charge neutrality of matter, $N_{p}=N_{e}$. Considering the quark structure of proton and neutron into account, we have $N_{u}=2N_{p}+N_{n}$, $N_{d}=N_{p}+2N_{n}$. Hence, one can write
\begin{equation}\label{eqn8}
\epsilon_{\alpha\beta} = \epsilon_{\alpha\beta}^e + (2+Y_{n})\epsilon_{\alpha\beta}^{u}+(1+2Y_{n})\epsilon_{\alpha\beta}^{d}, \hspace{2mm} Y_{n}=N_{n}/N_{e}.
\end{equation}

In our analysis, the PMNS matrix is considered to be different from its usual parameterization by a factor $P$=Diag$(e^{i\delta},1,1)$. The modified mixing matrix $U_v = P U P^{\ast}$, can be expressed as
\begin{widetext}
	\begin{equation}\label{eqn9}
	U_v(\theta_{12},\theta_{23},\theta_{13},\delta)=\begin{pmatrix}
	c_{12}c_{13} & s_{12}c_{13}e^{i\delta} & s_{13}\\
	-s_{12}c_{23}e^{-i\delta}-c_{12}s_{13}s_{23} & c_{12}c_{23}-s_{12}s_{13}s_{23}e^{i\delta} & c_{13}s_{23}\\
	s_{12}s_{23}e^{-i\delta}-c_{12}s_{13}c_{23} & -c_{12}s_{23}-s_{12}s_{13}c_{23}e^{i\delta} & c_{13}c_{23}
	\end{pmatrix},
	\end{equation}
\end{widetext}
where $c_{ij}=cos\theta_{ij}, s_{ij}=sin\theta_{ij}$ and $\delta$ is the $CP$ violating phase.
Due to the consideration of complex NSI parameters, there appears extra phase factor $\phi_{\alpha\beta}$ which can affect the correct estimation of $\delta$. To get rid of this difficulty PMNS matrix is specifically chosen as given in Eq. (\ref{eqn9}). This has been discussed in detail in \cite{Esteban3}. Another difficulty arises due to CPT symmetry under which the vacuum Hamiltonian has to transform as, $\mathcal{H}_{vac}\rightarrow-\mathcal{H}_{vac}^{\ast}$. As a consequence, the mass ordering $\Delta m_{31}^{2}$ gets reversed and the octant of $\theta_{12}$ is shifted from $0<\theta_{12}<\pi/4$ to $\pi/4<\theta_{12}<\pi/2$. To restore the CPT invariance of the neutrino oscillation probability in the presence of NSI, the following transformations are to be made simultaneously,
\begin{equation}
\begin{aligned}
\sin\theta_{12} \leftrightarrow  \cos\theta_{12}\,,\\
\Delta m_{31}^{2} \rightarrow -\Delta m_{31}^{2}+\Delta m_{21}^{2}\,,\\
\delta \rightarrow \pi-\delta\,,\\
\epsilon_{ee}-\epsilon_{\mu \mu}\rightarrow -(\epsilon_{ee}-\epsilon_{\mu \mu})-2\,,\\
\epsilon_{\tau \tau}-\epsilon_{\mu \mu}\rightarrow -(\epsilon_{\tau \tau}-\epsilon_{\mu \mu})\,,\\
\epsilon_{\alpha \beta}\rightarrow -\epsilon_{\alpha \beta}^{*}\,.
\end{aligned}
\end{equation}

From the global analysis including both oscillation and COHERENT data as shown in \cite{Esteban3}, we are left with two degenerate solutions: (i) LMA-Light solution ($\theta_{12} \approx 34^o$) with small NSI values and (ii) LMA-Dark octant ($\pi/4 < \theta_{12} < \pi/2$) with large values of NSI parameters. 
Oscillation data alone cannot lift this degeneracy. Hence non-oscillatory experiments such as COHERENT are useful to constraint NSI parameters \cite{Esteban:2018ppq}. In a recent analysis of COHERENT experiment including time and energy information \cite{Coloma:2019mbs}, it has been shown that LMA-Dark solution is discarded for a broad range of NSI parameters where the mediator mass is above $\sim \mathcal{O}(10)$ MeV. However, a few models have been constructed where a mediator of mass $\sim 10$ MeV is able to produce sufficiently large NSI \cite{Farzan:2017xzy,Denton:2018xmq,Farzan:2015doa}.

\subsection{Quantum correlation quantities}\label{QC}
Here we will briefly discuss some of the spatial as well as temporal quantum correlation measures used in this work.

{\it Flavour Entropy}: In classical information theory the {\it Shannon entropy} generally measures the uncertainty in the state of the physical system. In other words, it quantifies the information gained by learning about the outcome attained by measuring a system. A quantum mechanical analogue of Shannon entropy is {\it von Neumann entropy}, defined as $S(\rho) = - \rho \log_2 \rho$ for a system represented by density matrix $\rho$. It is zero for pure states and can attain its maximum value, $\log$ d, for a d-dimensional mixed state. If the compound system is pure, such as neutrinos, a standard measure of entanglement for a multipartite system can be defined as the sum of the von Neumann entropy of the reduced density matrix obtained by taking the trace over each one of the subsystems involved. Moreover, this measure can be considered as an {\it absolute} entanglement measure for a tripartite system since its nonzero value ensures the existence of the nonzero entanglement at least in one bipartition. 
	For the three flavor neutrino oscillation system we name it {\it flavour entanglement entropy} and can write it as a concave function of transition probabilities \cite{Blasone:2007vw,Dixit:2018kev},
\begin{align}
S\left(|{U_f}_{ij}|^{2}\right) =& - \sum\limits_{j=1}^3|{U_f}_{ij}|^{2} \log_{2}\left(|{U_f}_{ij}|^{2}\right)\\& - \sum\limits_{j=1}^3\left(1-|{U_f}_{ij}|^{2}\right) \log_{2}\left(1-|{U_f}_{ij}|^{2}\right)\,, \label{eqn10}
\end{align}
where $i = 1,2,3$ corresponding to initial neutrino flavour $\alpha = e,\mu,\tau$, respectively and $U_f$ is the evolution operator for neutrino system. Minimum value of $S=0$, {\it i.e.,} no entanglement condition is obtained if any one of $P_{\alpha\beta}=1$ and the maximum value or upper bound, $S=2.75$, of this parameter can be approached when $P_{\mu e} = P_{e \mu} = P_{\mu \tau} = \frac{1}{3}$, {\it i.e.,} all the three flavors are equally probable.\\

{\it Genuine Tripartite Entanglement}: Another measure of tripartite entanglement, in the {\it genuine} sense, can be defined as cube of the geometric mean of von Neumann entropies of each bipartite section and can be expressed as following \cite{Dixit:2018kev}
\begin{equation}\label{eqn11}
G\left(|{U_f}_{ij}|^{2}\right) = \Pi_{j=1,2,3}H\left(|{U_f}_{ij}|^{2}\right),
\end{equation}
where $i = 1,2,3$ corresponding to $\alpha = e,\mu,\tau$ and $H(x)=-x \log_{2}(x) - (1-x) \log_{2}(1-x)$. Here, $G$ is called a measure of genuine entanglement since the nonzero value of this measure can be obtained only when all the subsystems are entangled with each other. $G$ will be zero if any of the subsystems is not entangled with the rest of the system.\\

{\it Leggett-Garg type Inequality (LGtI)}: The above two measures of entanglement can be considered as measures of correlations between spatially separated systems. Leggett-Garg inequalities (LGI), based on the assumptions of (i) {\it macro-realism (MR), i.e.,} a macroscopic system with two or more macroscopically distinct states available to it will always be available in one of those states, and (ii) {\it noninvasive measurement (NIM), i.e.,} it is possible to perform a measurement on a system without even disturbing its dynamics, capture the correlations among measurements performed on a system at different times. Mainly, LGIs were introduced to manifest macroscopic coherence which means that up to what level quantum mechanics is applied on a many-particle system exhibiting decoherence \cite{Leggett:1985zz}. On the other hand, LGI tests also give space to test the notion of realism which introduces the concept of hidden-variable theories and implies that a physical system posses predefined values of all of its parameters independent of measurement \cite{Huelga2,Emary}. Therefore, the violation of these inequalities will indicate that such hidden-variable theory cannot be considered as an alternative to describe the time evolution of a quantum mechanical system. 

The LGI parameter is basically a linear combination of autocorrelation functions $C(t_i,t_j) = \frac{1}{2} Tr[\{\hat{Q}(t_i),\hat{Q}(t_j)\} \rho(t_0)]$ with $\rho(t_0)$ being the initial state of a given system at time $t = 0$  and can be written as \cite{Leggett:1985zz}
\begin{equation}
K_3 = C(t_1,t_2) + C(t_2,t_3) - C(t_1,t_3) \leq 1.
\end{equation}

Here, $\hat{Q}$ is a dichotomic observable, {\it i.e.,} $\hat{Q} = \pm 1$ with $\hat{Q} = +1$ if the system is found in the target state and $\hat{Q} = -1$ otherwise.
Measurement of neutrinos destroys the  $NIM$ assumption. Hence the weaker condition of {\it stationarity} is applied to relax this assumption \cite{Huelga}. Due to the stationarity condition, functions $C(t_i,t_j)$ now depend only on the time difference $t_j - t_i$. The $K_3$ quantity can be written as \cite{Formaggio:2016cuh,Naikoo:2017fos,Naikoo:2019eec}
\begin{equation}
K_3 = 2 C(0,t) - C(0,2t) \leq 1, \label{LGtI}
\end{equation}
for $t_1 = 0$ and $t_2 - t_1 = t_3 - t_2 \equiv t$. 
\begin{equation}
K_{3}=1 + 2P_{\alpha \beta}(2L,E) - 4P_{\alpha \beta}(L,E).
\end{equation}\label{eqn12}
Here we have applied the condition $t \equiv L$ for ultrarelativistic neutrinos. It was shown in \cite{Formaggio:2016cuh} that the parameter $K_{3}$ can be determined experimentally by making use of the condition  $P_{\alpha \beta}(2L,E) = P_{\alpha \beta}(L,\tilde{E})$  by suitable choice of $E$ and $\tilde{E}$.

\section{Results and discussion} \label{Results}
In this section, we analyze various measures of quantum correlations present in the neutrino system for the SM and NSI interactions. We present our results for the DUNE experiment set up. The values of NSI parameters (within 1$\sigma$ interval) have been extracted from the global analysis of neutrino oscillation and coherent neutrino scattering COHERENT experimental data as given in Table \ref{Tab2}. Here, the effect of parameter $\epsilon_{\alpha\beta}^e$ is neglected \cite{Esteban3}. We also make use of the values of standard neutrino oscillation parameters from a recent analysis \cite{deSalas:2020pgw} which is given in Table \ref{Tab}. The matter density potential, $\rho$, is taken to be 2.8 gm/cc which is appropriate for the DUNE experiment. 

\begin{figure*}[t]
	\begin{center}
		\includegraphics[width=80mm]{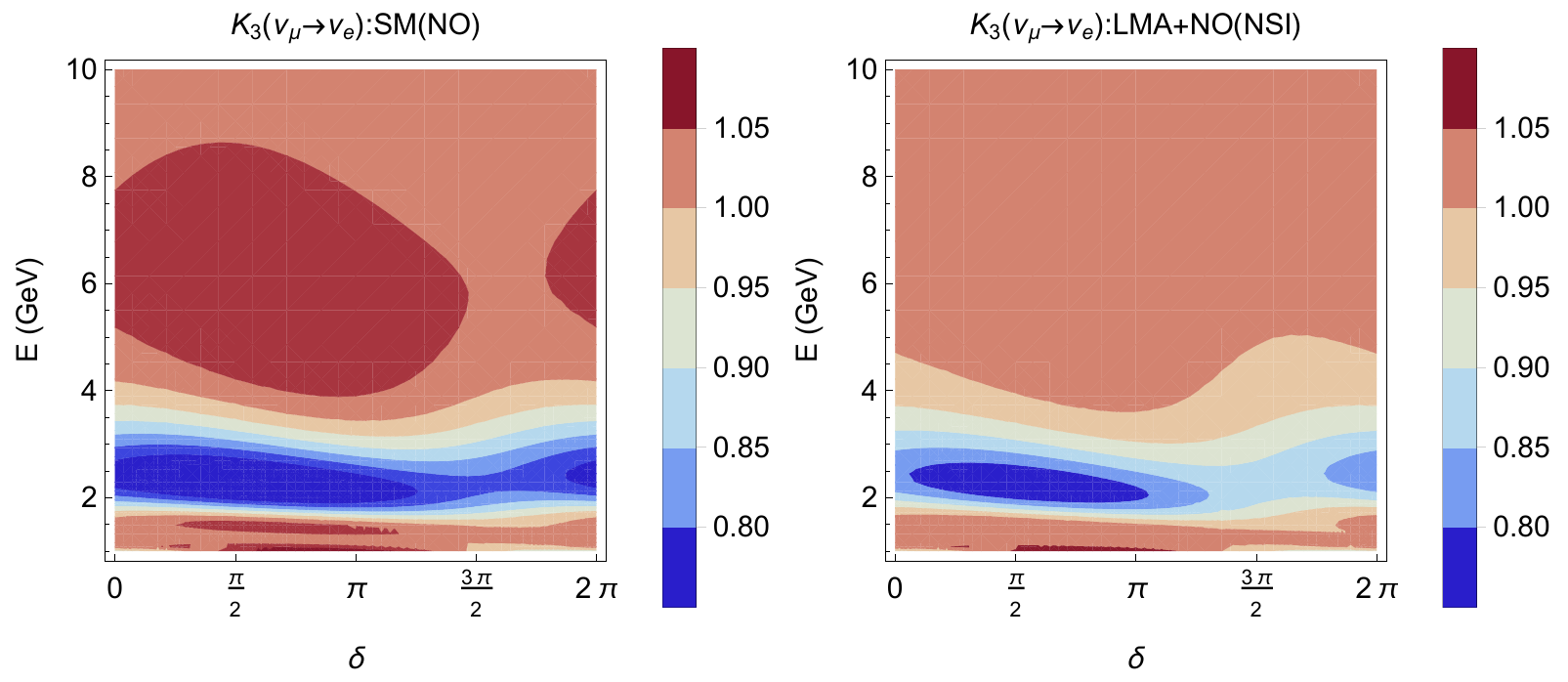}
		\includegraphics[width=40mm]{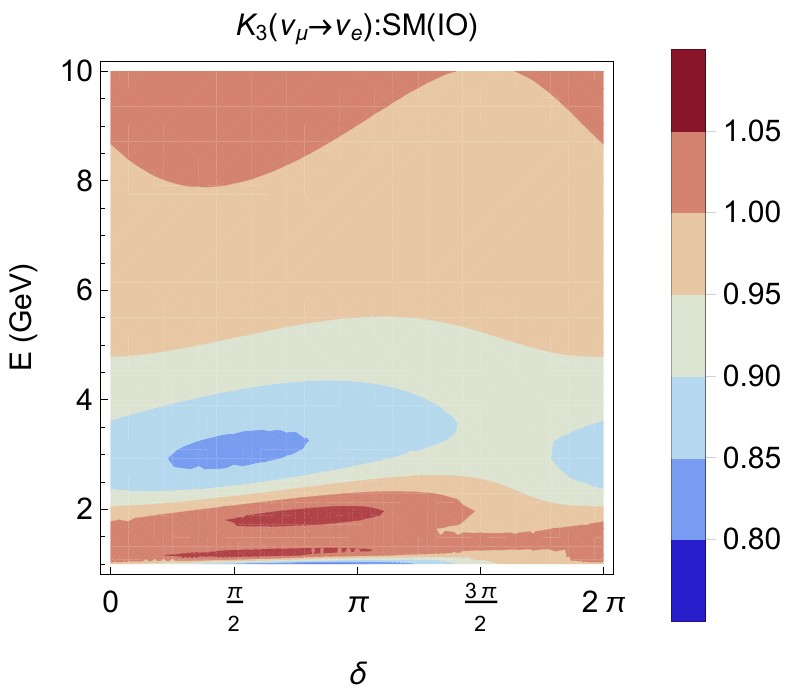}
		\includegraphics[width=40mm]{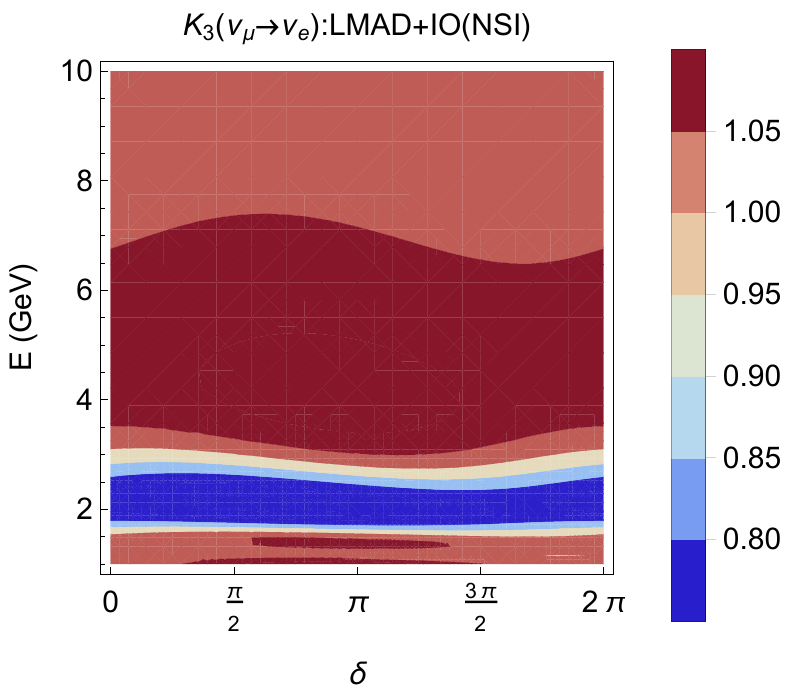}
		\caption{$K_3$ has been plotted in the $E-\delta$ plane in case of SM-interaction with NO (first), LMA-Light + NO (second), SM + IO (third) and LMA-Dark + IO scenario (fourth) in the context of DUNE experiment ($L = 1300$ km, $E = 1 - 10$ GeV).}
		\label{LGI}
	\end{center}
\end{figure*}

We first discuss the behaviour of temporal correlations under the influence of NSI and SM interactions as portrayed in Fig. \ref{LGI}. The results are summarized in Table \ref{Tab3}. Since, our aim is to explore the signatures of new physics, hence, in Table \ref{Tab3} we have provided specific ranges of neutrino-energy and $\delta_{CP}$ where one can significantly distinguish the effects of NSI and SM interaction. Later, we also perform a similar analysis for spatial correlation measures. 

\begin{table}
	
	\begin{center} 
		
		\begin{tabular}{|p{2.5cm}|p{2.5cm}|p{2.5cm}|}
			
			\hline
			
			Parameters & LMA-Light +NO \par ($\sim 1\sigma$ interval)  & LMA-Dark +IO \par ( $\sim 1\sigma$ interval) \\ \hline\hline
			$\epsilon_{ee}-\epsilon_{\mu\mu}$ & [-0.5, 0.25] & [-2.5, -1.75] \\ \hline
			$\epsilon_{\tau\tau}-\epsilon_{\mu\mu}$ &  [0, 0.1]  &  [-0.2,0]\\ \hline
			$|\epsilon_{e\mu}|$  &  [0, 0.1]  &  [0, 0.1]\\ \hline
			$|\epsilon_{e\tau}|$  &  [0, 0.75]  &  [0,0.25]\\ \hline
			$|\epsilon_{\mu\tau}|$  &  [0, 0.02]  &  [0, 0.025]\\ \hline
			$\phi_{e\mu}$  &  [$67.5^{o}, 281.25^{o}$]  &  [$0^{o}, 90^{o}$], [$247.5^{o}, 360^{o}$]\\ \hline
			$\phi_{e\tau}$ &  [$0^{o},360^{o}$] &  [$0^{o},360^{o}$]\\ \hline
			$\phi_{\mu\tau}$ & [$0^{o},360^{o}$]  &  [$0^{o},360^{o}$]\\ \hline
			$\delta$  & [$180^{o}$, $315^{o}$] &  [$213.75^{o}$, $360^{o}$]\\
			\hline
		\end{tabular}
		\caption{1$\sigma$ interval of NSI parameters taken from Ref. \cite{Esteban3}.}
		\label{Tab2}
	\end{center}
\end{table}

\begin{table}
	\begin{center}
		\begin{tabular}{|c       c |} 
			\hline
			Parameters & Best fit $\pm 1\sigma$ \\ [0.7ex] 
			\hline\hline
			$\theta_{12}^o$  &   $34.3{\pm 1.0}$ \\
			\hline
			$\Delta m_{21}^2 \times 10^{-5} eV^2$ & $7.5^{+0.22}_{-0.20}$ \\
			\hline
			$\theta_{23}^o$(NO) & $48.79^{+0.93}_{-1.25}$ \\
			\hline
			$\theta_{13}^o$(NO) & $8.58^{+0.11}_{-0.15}$\\
			\hline
			$|\Delta m_{31}^2| \times 10^{-3}eV^{2}$ (NO) & $2.56^{+0.03}_{-0.04}$ \\
			\hline
			\hline
			$\theta_{23}^{o}$(IO) & $48.79^{+1.04}_{-1.30}$ \\
			\hline
			$\theta_{13}^{o}$(IO) & $8.63^{+0.11}_{-0.15}$\\
			\hline
			$|\Delta m_{31}^2| \times 10^{-3}eV^2$(IO) & $2.46 \pm 0.03$ \\
			\hline
		\end{tabular}
		\caption{Standard neutrino oscillation parameters with 1$\sigma$ intervals obtained in \cite{deSalas:2020pgw}.}
		\label{Tab}
	\end{center}
\end{table}
\FloatBarrier

{\it Temporal correlation:} The behaviour of $K_3$ for three flavour neutrino oscillation scenario is depicted in Fig. \ref{LGI} in case of SM interaction and NSI effect for Dune experimental setup ($L = 1300$ km, $E = 1 - 10$ GeV). In Table \ref{Tab3} we have indicated peculiar ranges of $\delta_{CP}$ for specific neutrino-energies where $K_3$ parameter exceeds its classical bound. It can be seen in Fig. \ref{LGI} and Table \ref{Tab3} that the parameter $K_3$ is sensitive to NSI and SM interaction for both normal and inverted mass ordering and is violated for almost entire energy spectra of 1 - 10 GeV. However, for certain energy values, $K_3$ violation is observable only for specific choices of possible solutions.
For example:
\begin{itemize}
	\item At $E \approx$ 3 GeV, $K_{3}$ can exceed the classical limit only for the LMA-Dark sector + IO scenario.
	\item At $E \approx$ 2 GeV, for $\pi/2\lesssim \delta \lesssim 23 \pi/16$, violation of $K_{3}$ is possible only for SM interaction (with IO).
\end{itemize}

\begin{table}
	\centering
	\setlength{\tabcolsep}{0.3\tabcolsep}
	\begin{tabular}{|c|c|c|c|c|}
		\hline			
		E in GeV & SM+NO & LMA-L+NO  &  SM+IO  &  LMA-D+IO\\
		\hline
		1.0 & 5$\pi$/16 - 23$\pi$/16 & 5$\pi$/16 - 11 $\pi$/8 & - & 0 - 2$\pi$\\ \hline
		1.5 & 0 - 21$\pi$/16 & 0 - 9$\pi$/8 & 0 - 2$\pi$ & 0 - 2$\pi$\\ \hline
		2.0 & - & - & $\pi$/2 - 23$\pi$/16 & - \\ \hline
		2.5 & - & - & - & - \\ \hline
		3.0 & - & - & - & 5$\pi$/8 - 27$\pi$/16 \\ \hline
		3.5 & 14$\pi$/16 - 5$\pi$/4 & - & - & 0 - 2$\pi$\\  \hline
		4.0 & $\pi$/4 - 25$\pi$/16 & 7$\pi$/16 - 5$\pi$/4 & - & 0 - 2$\pi$ \\ \hline
		4.5 & 0 - 2$\pi$ & $\pi$/8 - 11$\pi$/8 &  - & 0 - 2$\pi$\\ \hline
		5.0 & 0 - 2$\pi$ & 0 - 2$\pi$ & - & 0 - 2$\pi$\\ \hline
		6.0 & 0 - 2$\pi$ & 0 - 2$\pi$ & - & 0 - 2$\pi$\\ \hline
		7.0 & 0 - 2$\pi$ & 0 - 2$\pi$ & - & 0 - 2$\pi$\\ \hline
		8.0 & 0 - 2$\pi$ & 0 - 2$\pi$ & 5$\pi$/8 & 0 - 2$\pi$\\ \hline
		9.0 & 0 - 2$\pi$ & 0 - 2$\pi$ & 0 - $\pi$ & 0 - 2$\pi$\\ \hline
		10.0 & 0 - 2$\pi$ & 0 - 2$\pi$ & 0-2$\pi$  & 0 - 2$\pi$\\
		\hline
	\end{tabular}
	\caption{Specific ranges of $\delta$ have been provided for distinct values of neutrino-energy where temporal correlation parameter $K_{3}$ exceeds the value 1, $i.e.,$ LGtI is violated.}
	\label{Tab3}
\end{table}

\begin{figure} [h] 
	\includegraphics[height=40mm,width=.66\columnwidth]{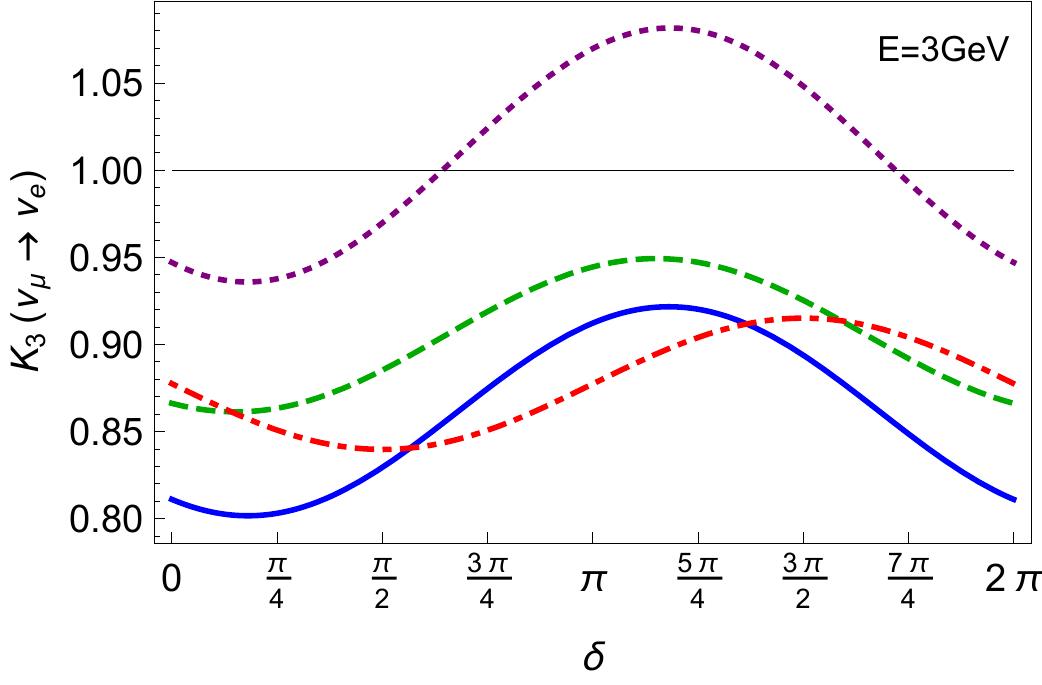} \\
	\includegraphics[height=40mm,width=.66\columnwidth]{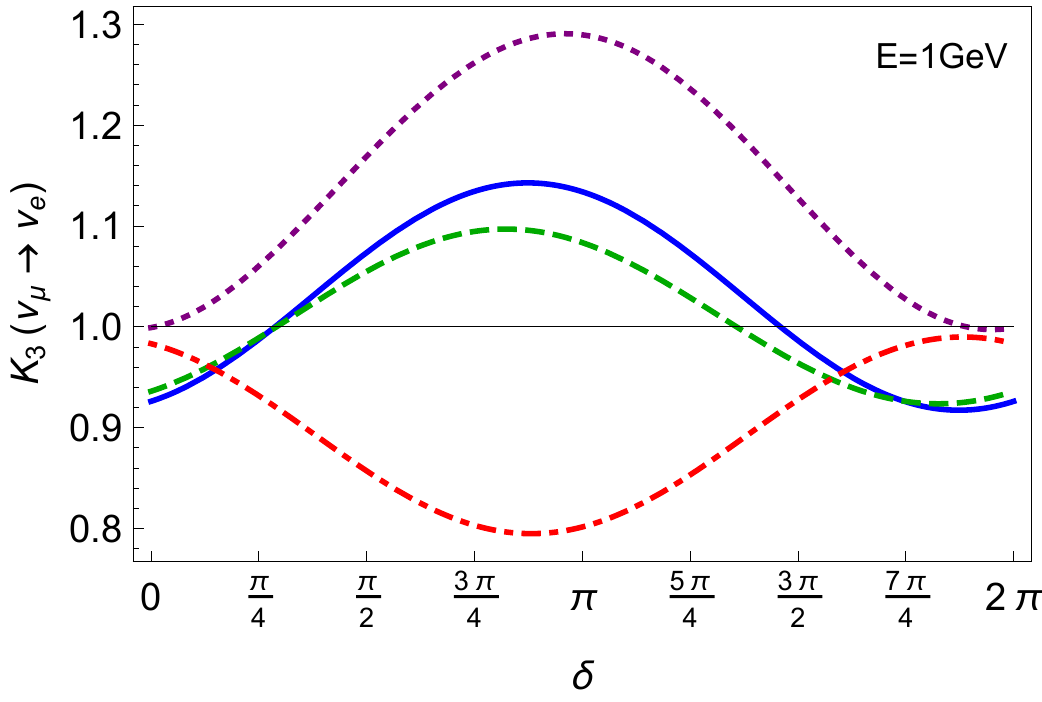} \\ 
	\hspace*{1.3cm}
	\includegraphics[height=40mm,width=.92\columnwidth]{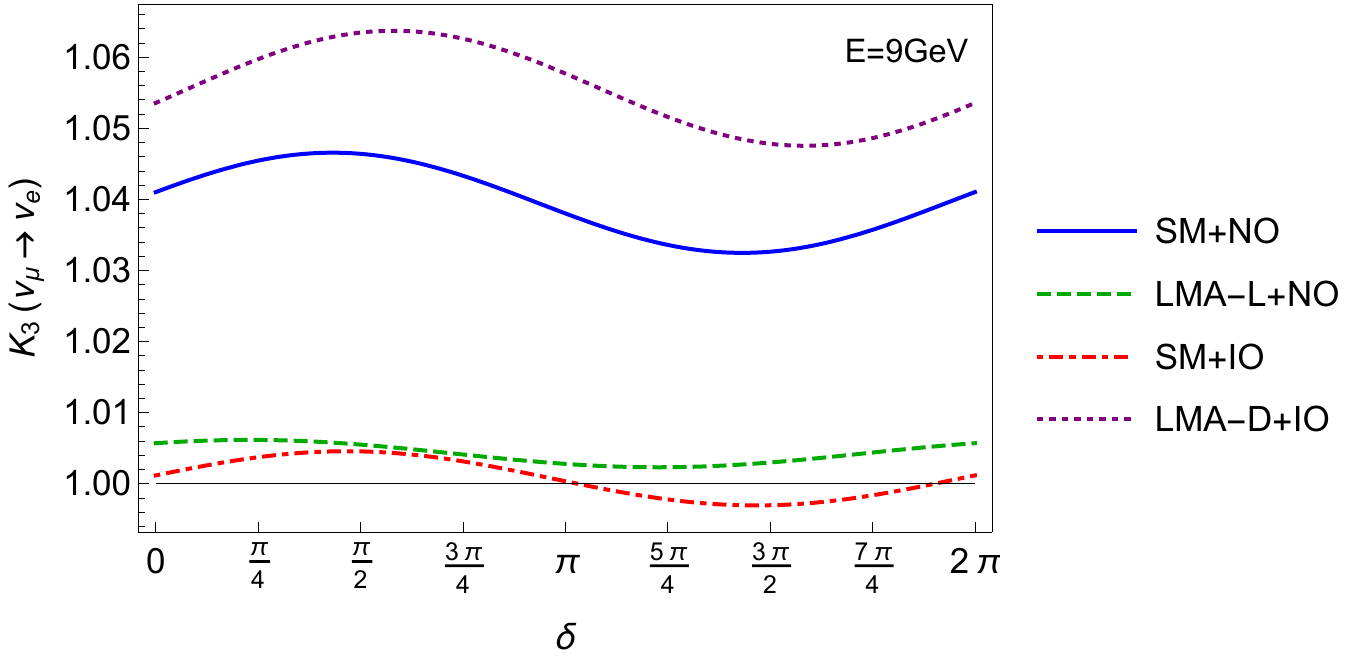}
	\caption{Parameter $K_3$ is plotted with respect to $\delta$ (radian) at $E = 1$ GeV (upper), $E = 3$ GeV (middle) and $E = 9$ GeV (lower) for the case of SM-interaction with NO (blue solid curves), LMA-Light + NO (green dashed), SM + IO (red dot-dashed) and LMA-Dark + IO scenario (purple dotted). These plots are obtained in the context of DUNE experimental setup ($i.e.,$ $L = 1300$ km).}
	\label{K3}
\end{figure}

\begin{figure*}[t]
	\begin{center}
		\includegraphics[width=40mm]{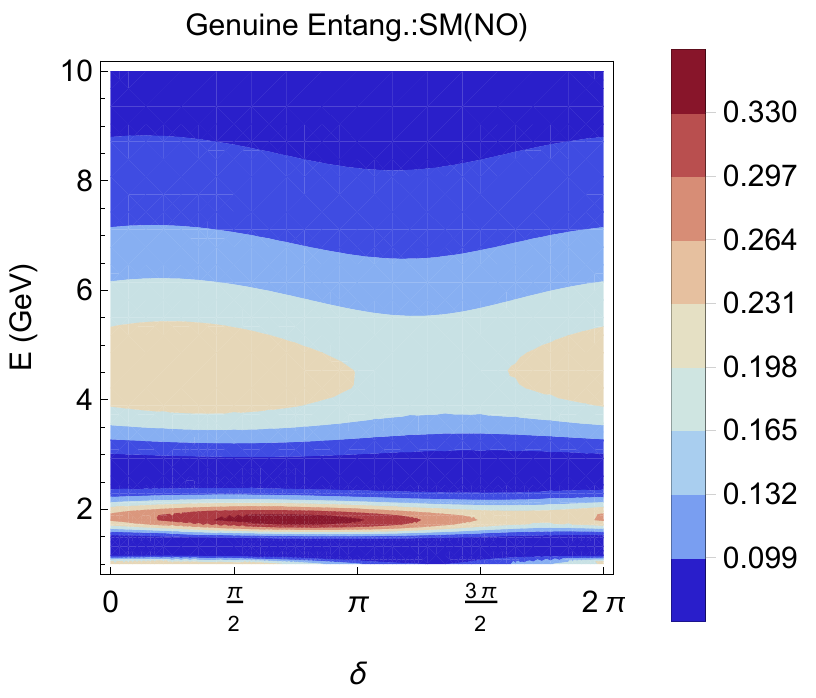}
		\includegraphics[width=40mm]{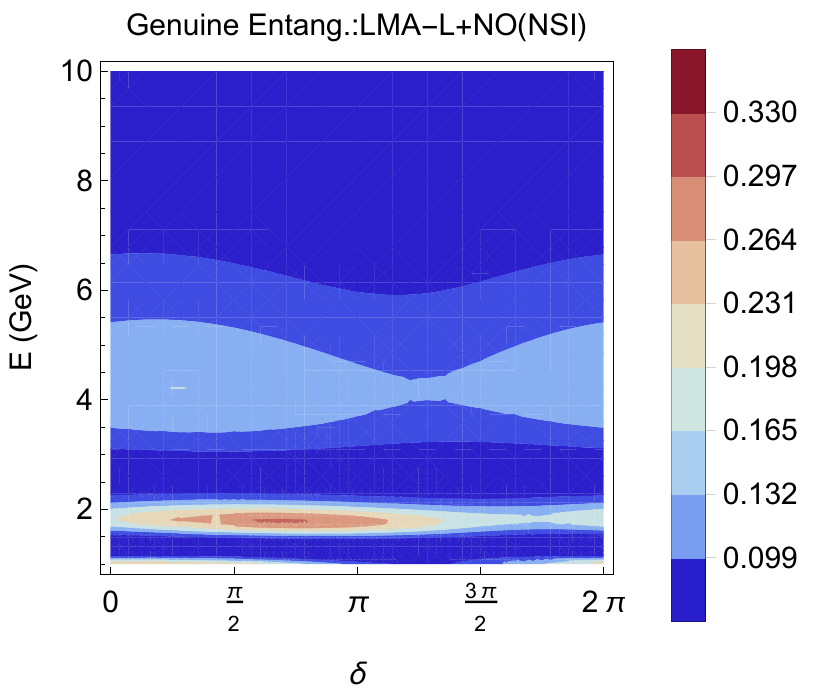}
		\includegraphics[width=40mm]{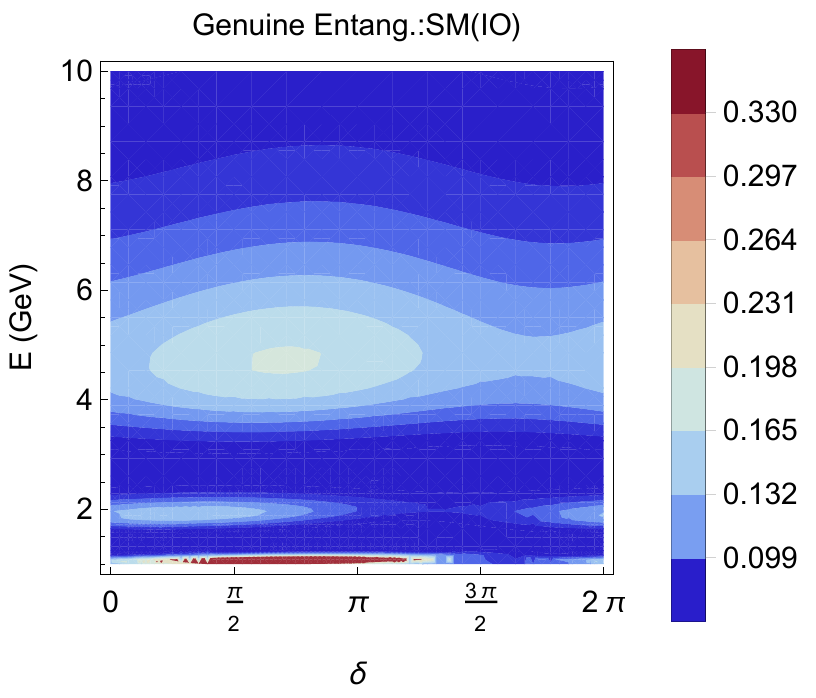}
		\includegraphics[width=40mm]{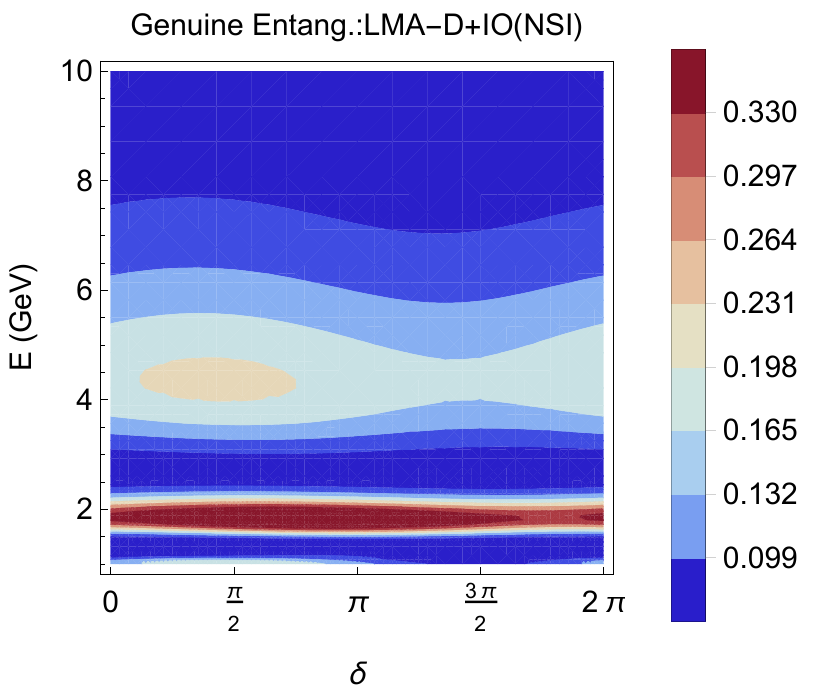}\\
		\includegraphics[width=80mm]{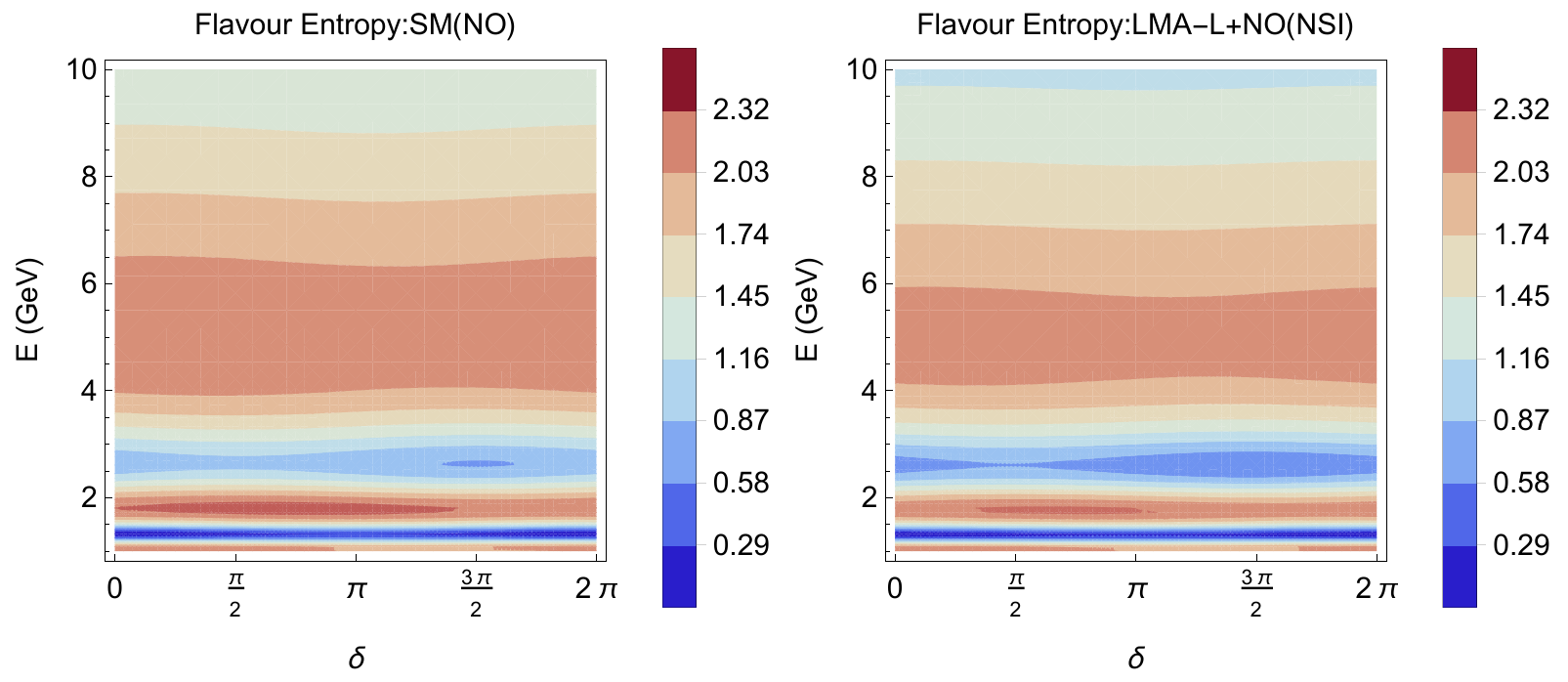}
		\includegraphics[width=40mm]{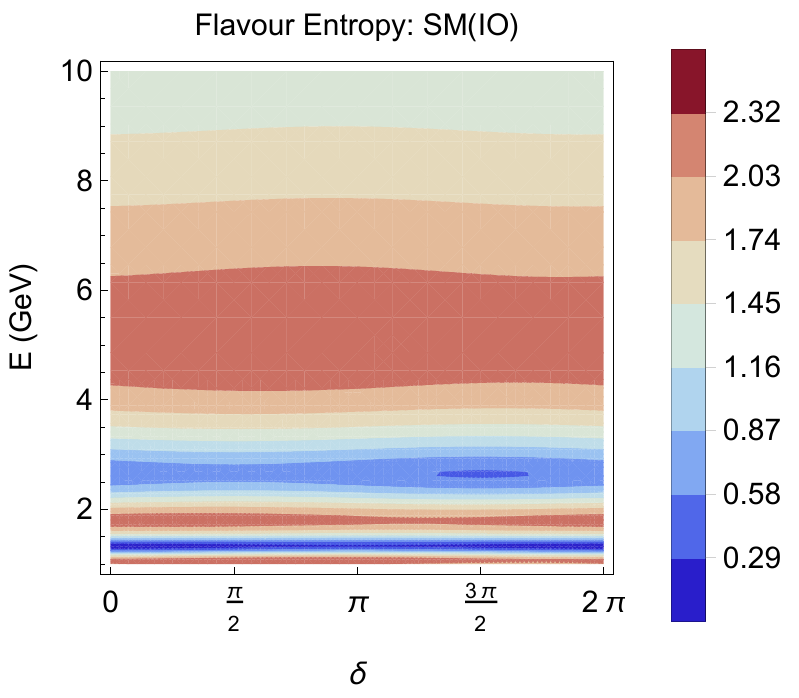}
		\includegraphics[width=40mm]{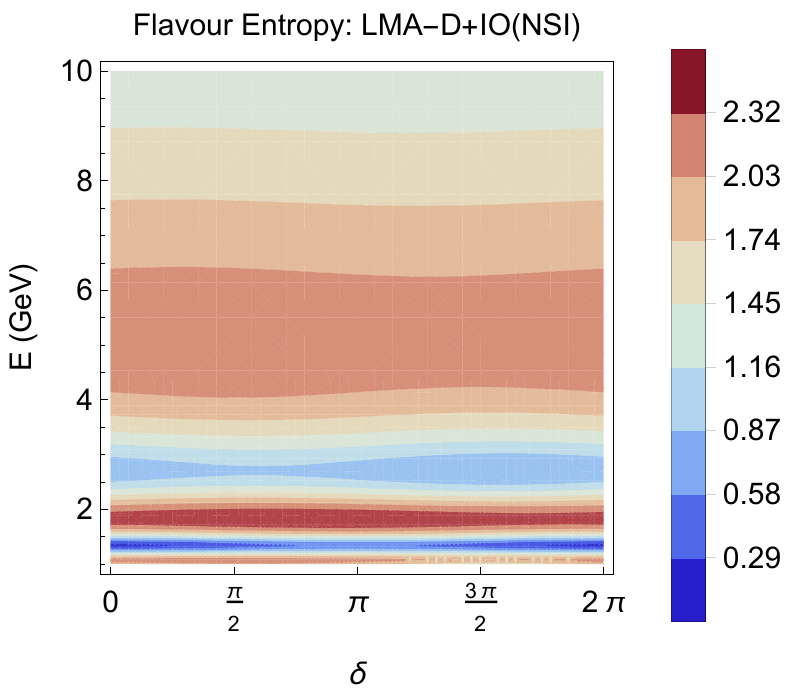}
		\caption{Genuine tripartite entanglement $G$ (upper figures) and flavor entropy $S$ (lower figures) have been plotted in the $E-\delta$ plane in case of SM-interaction with NO (first column), LMA-Light + NO (second column), SM + IO (third column) and LMA-Dark + IO scenario (fourth column) in the context of DUNE experiment ($L = 1300$ km, $E = 1 - 10$ GeV).}
		\label{QC_LMA}
	\end{center}
\end{figure*}

\begin{figure*}
	\begin{center}
		\includegraphics[width=48mm]{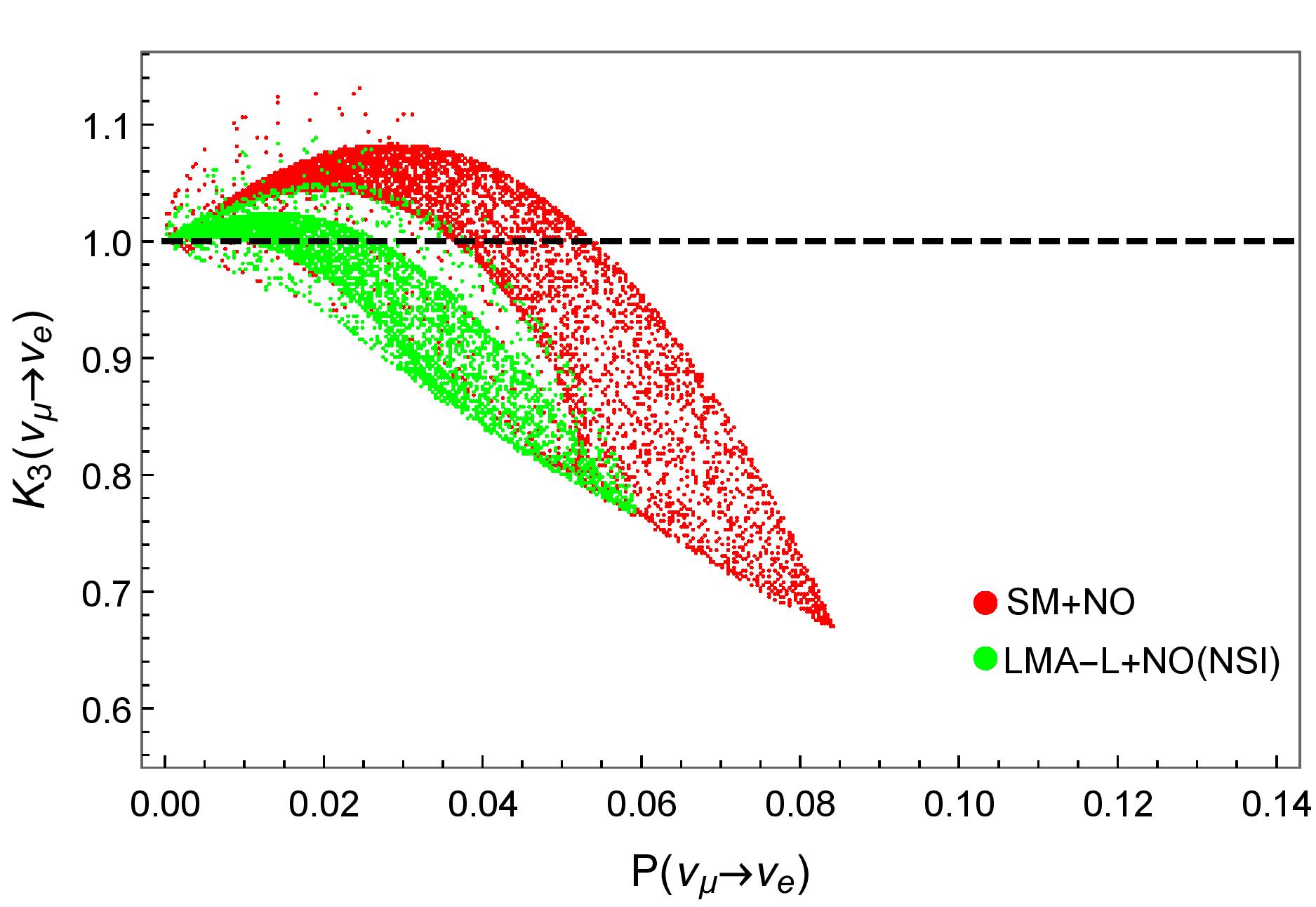}
		\includegraphics[width=48mm]{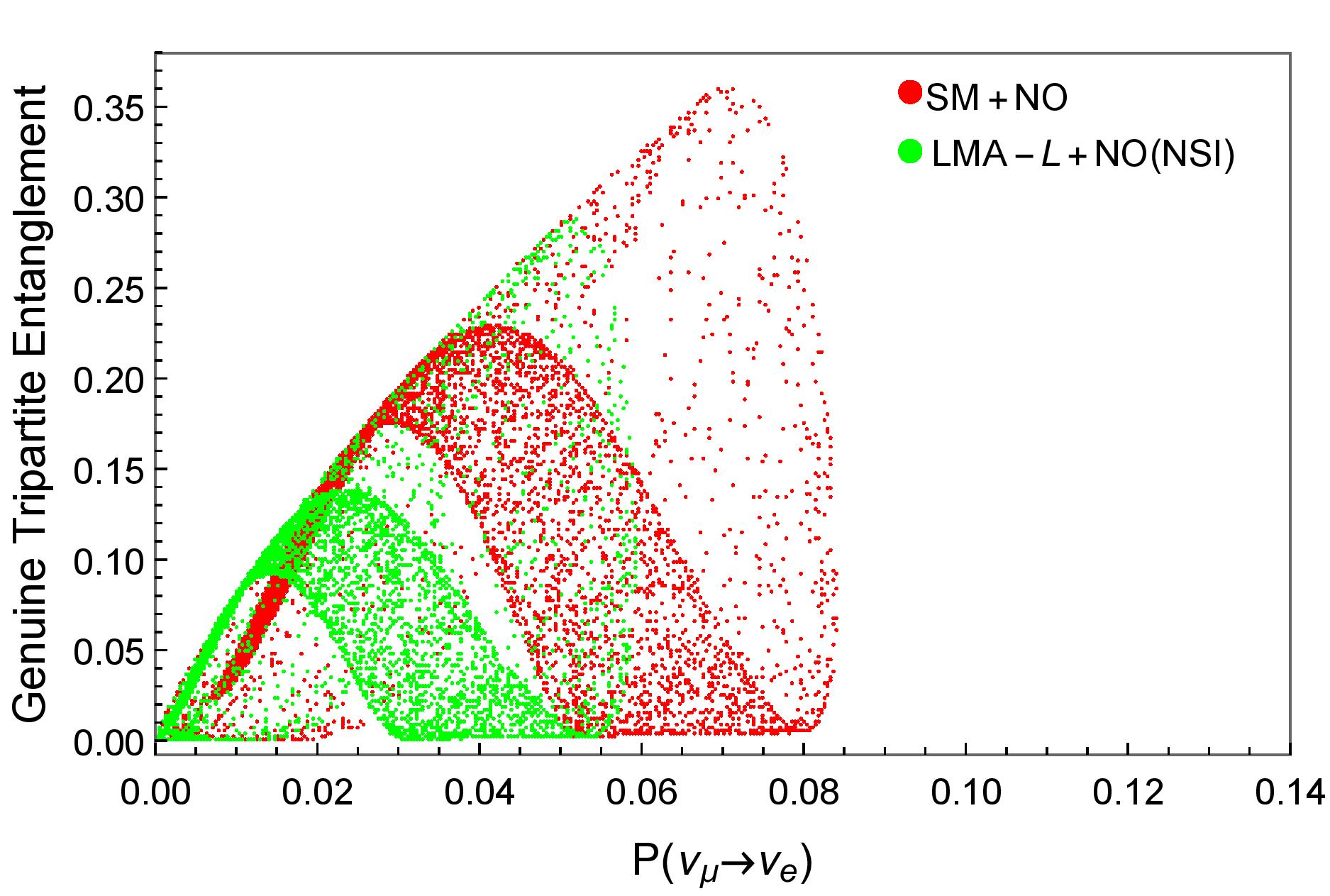}
		\includegraphics[width=48mm]{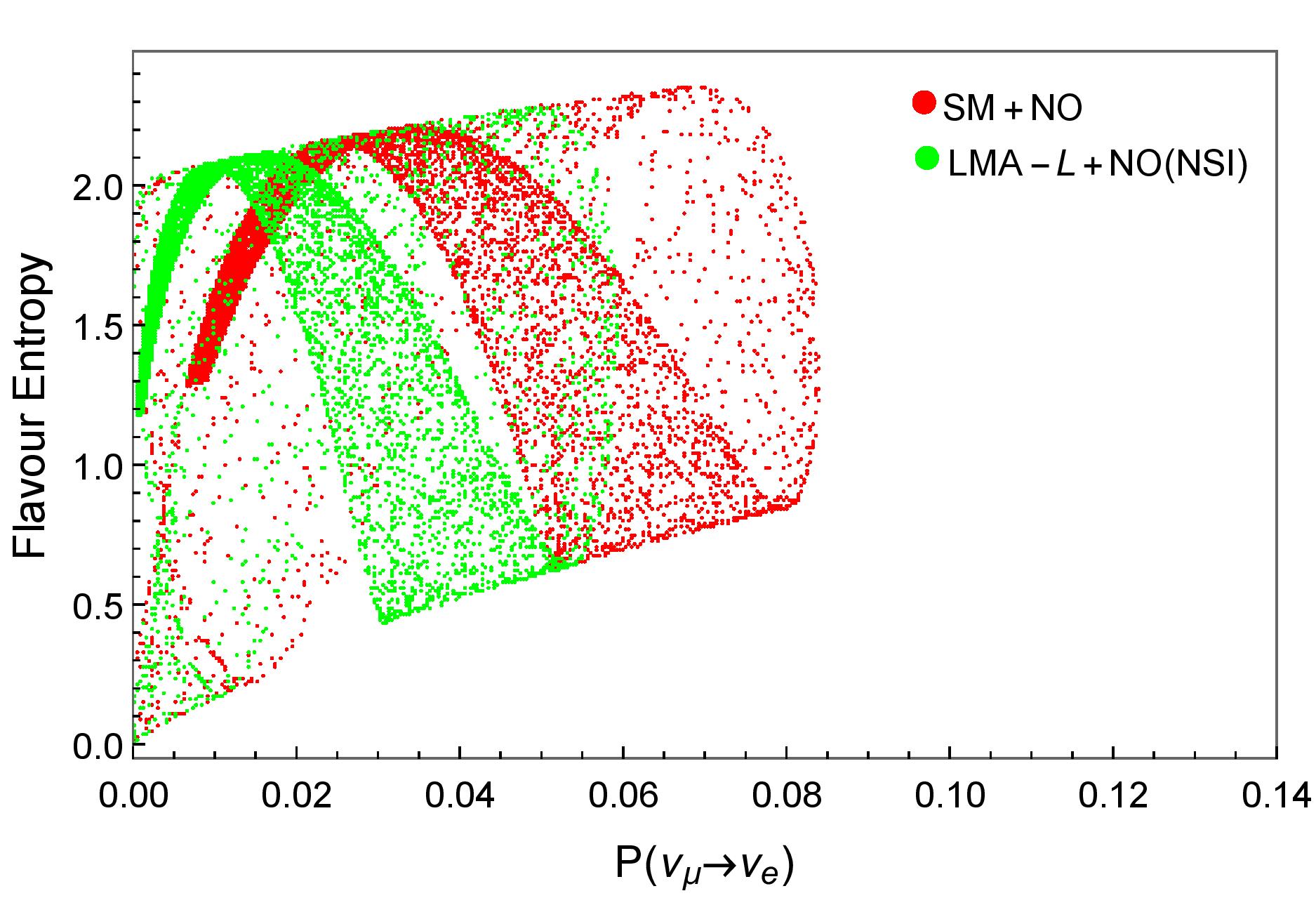}\\
		\includegraphics[width=48mm]{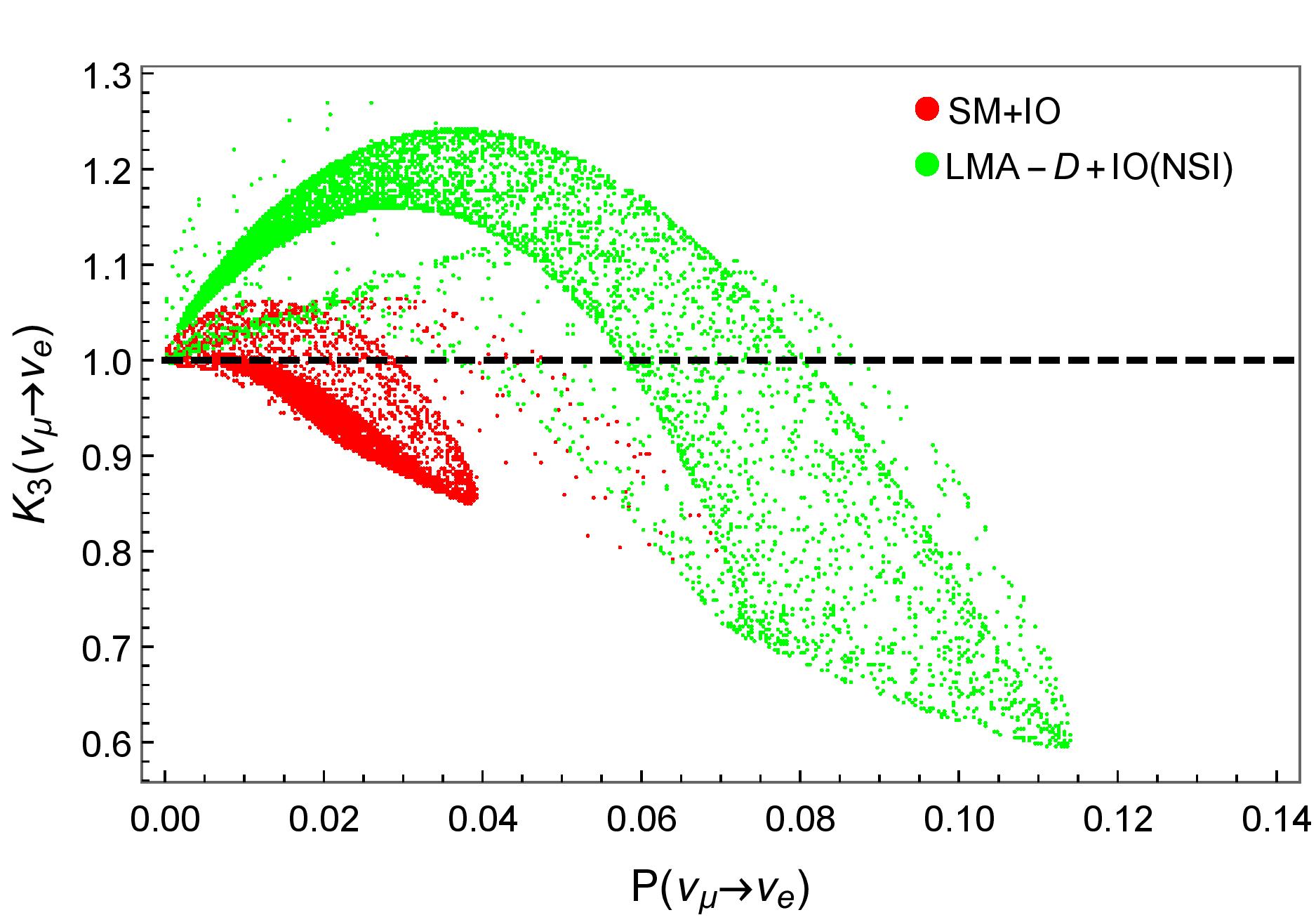}
		\includegraphics[width=48mm]{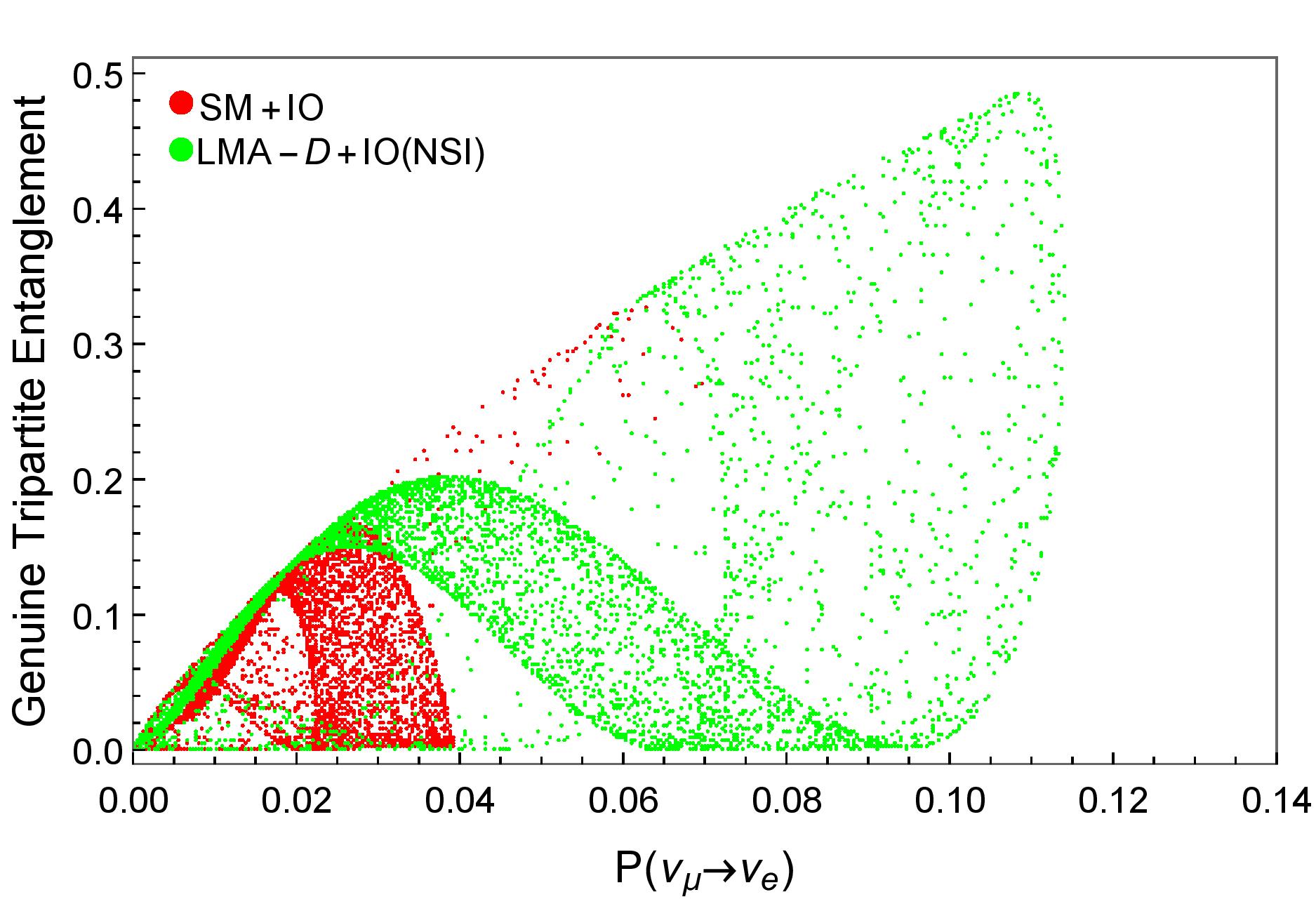}
		\includegraphics[width=48mm]{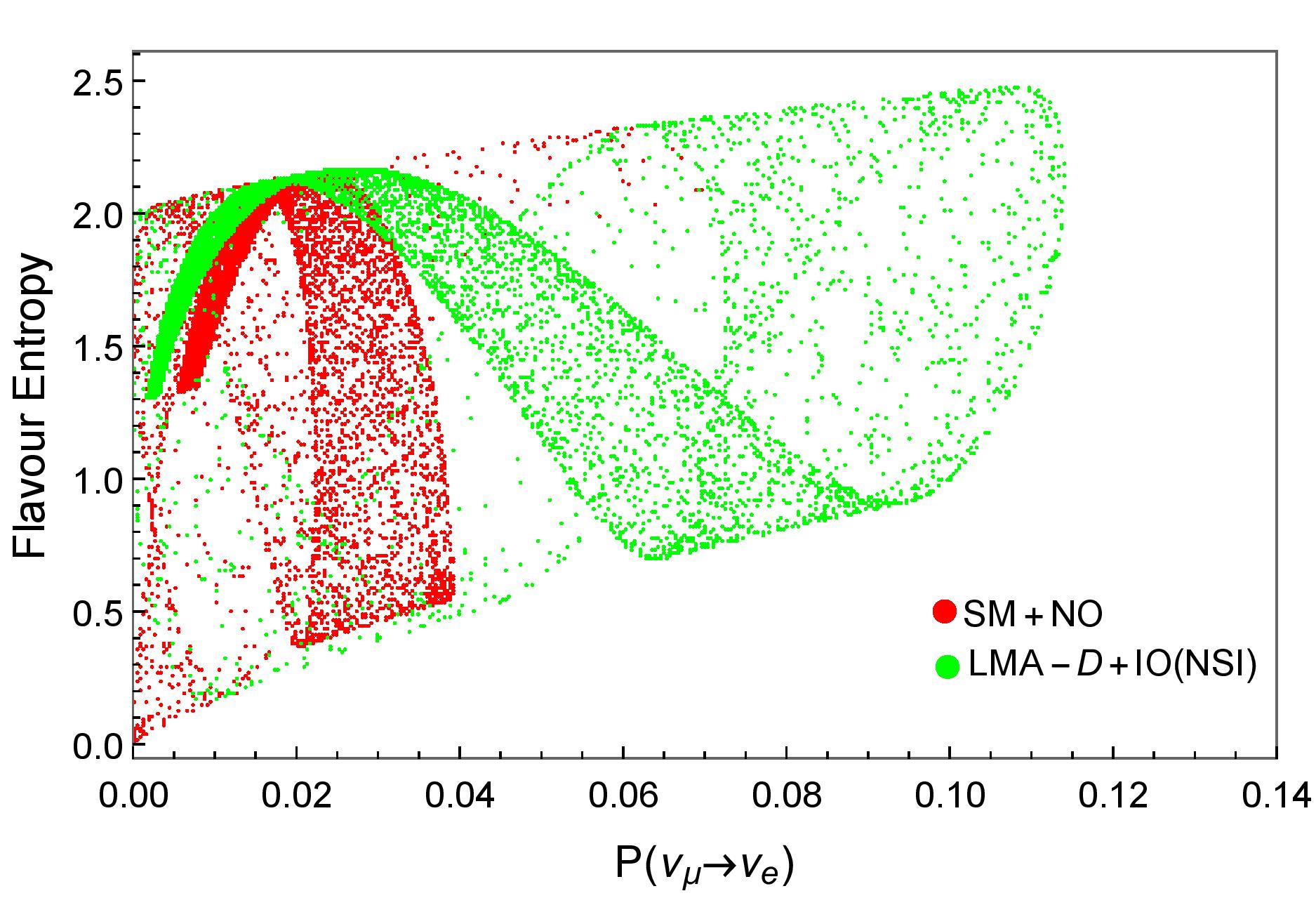}
		\caption{This figure represents the correlation of nonclassicality measure $K_3$ (first column), genuine tripartite entanglement $G$ (second column) and flavor entropy $S$ (third column) with transition probability $P(\nu_{\mu} \rightarrow \nu_{e})$.}
		\label{correlationPlot}
	\end{center}
\end{figure*}

Moreover, it can be seen for the lower energies ($E \approx 1 - 1.5$ GeV) and higher ($E \approx 8 - 10$ GeV) energy values, it is rather difficult to distinguish the effects of SM interaction and NSI. These facts are illustrated in Fig. \ref{K3} where $K_{3}$ is plotted with $\delta$ for $E = 3$ GeV (upper), 1 GeV (middle) and 9 GeV (lower) considering the $\nu_{\mu} \rightarrow \nu_e$ channel. The implementation of LGtI test requires measurements of neutrino transition probability at multiple spatial positions. For e.g, in order to measure the LGtI parameter $K_3$, one needs to measure $\nu_{\mu} \to \nu_{e}$ transition probability at two distinct spatial positions $L$ and $2L$ which is not possible as the baseline is fixed in the current neutrino experimental facilities. However, the measurements at $L$ and $2L$ for the fixed energy E can be translated to the measurements performed for two distinct energy values E and $\tilde{E}$ satisfying the relation $P(E,2L) = P(\tilde{E},L)$ for a fixed baseline $L$ \cite{Formaggio:2016cuh}. Therefore one can implement the LGtI tests using data obtained in a broad neutrino-energy spectrum. However from the entire dataset only those paired measurements can be utilized for which $P(E,2L) = P(\tilde{E},L)$. Hence measurements with very high statistics would be required to establish a clear signature of new physics through LGtI violations at DUNE.

{\it Spatial correlations:} The prediction for genuine tripartite entanglement for different scenarios are shown in the upper panel of Fig. \ref{QC_LMA}. It is found that the NSI (LMA-Light + NO) appears to suppress the value of entanglement in comparison with the entanglement in case of SM for normal mass ordering. It is seen that in case of SM + NO, large value of entanglement ($> 0.23$) is attained for $4 ~{\rm GeV} \leq E \leq 5.5$ GeV with $0 \leq \delta \leq \pi$ and $7\pi/4 \leq \delta \leq 2\pi$. On the other hand, in the same energy interval, the entanglement is reduced due to LMA-Light + NO and remains in the range $0.13 \lesssim G \lesssim 0.16$ for all values of $\delta$. In case of SM interaction, maximum value of G ($\geq 0.33$) is approached, while with NSI (LMA-Light + NO) effects, such a large value is not allowed.

In case of SM interaction with IO, the genuine entanglement attains quite low value ($\lesssim 0.19$) which is evident from the plot in the third column of upper panel of Fig. \ref{QC_LMA} (except for a very narrow region at E $\sim 1$ GeV, where $G > 0.33$). Conversely, NSI (LMA-Dark + IO) effects, enhances $G$ up to $\sim 0.26$ in the range $4~{\rm GeV} \leq E \leq 5 ~{\rm GeV}$ 
with a large amplification ($G > 0.33$) around 2 GeV for all values of $\delta$.

Further, it can be seen in the lower panel of Fig. \ref{QC_LMA}, that the variations of flavour entropy in the $E-\delta$ plane are almost similar for all the four scenarios (SM \& NSI with NO \& IO). The maximum value of flavor entropy is always $\approx 2.32$. This indicates that unlike genuine entanglement, the effect of NSI on the residual entanglement is extremely small.

Since DUNE aims to observe $\nu_{\mu}\rightarrow\nu_{e}$ oscillation channel \cite{Acciarri:2016crz}, we presented the correlation of $K_3$, $G$ and $S$ parameters with probability $P(\nu_{\mu} \rightarrow \nu_{e}$) in Fig. \ref{correlationPlot}. The purpose of these plots is to show the correlation of distinct nonclassicality measures with experimentally observable quantities in neutrino oscillations. Here red and green dots illustrate respectively, the cases of SM and NSI interactions. It is clear that in case of NO, the maximum value of probability $P(\nu_{\mu} \rightarrow \nu_{e})$ can be $\sim$ 0.08 for the SM interaction, while NSI (LMA-Light) sector can suppress this probability up to $\approx 0.06$. Similarly, violation of LGtI will be enhanced in case of SM interaction ({\it i.e.,} the value of $K_3$ is $\gtrsim 1.1$), whilst, NSI (LMA-Light) decreases this violation up to $K_3^{max} \sim 1.08$. It is interesting to note that the violation of $K_{3}$ at probability $P \sim 0.05$ can be induced only due to SM interaction in case of normal mass ordering.

For IO, LMA-Dark sector of parameters enhances the value of $P(\nu_{\mu} \rightarrow \nu_{e})$ up to $\sim 0.11$ and also increases the violation of LGtI, $i.e.,$ $K_3^{max} \sim 1.22$. While SM prefers a relatively lower value of both $P(\nu_{\mu}  \rightarrow \nu_{e})$ and $K_3$, it can be seen from the plot that if $K_3$ is violated for $P(\nu_{\mu} \rightarrow \nu_{e}) > 0.07$, then it can be due to LMA-Dark solution only.

Similar features are observed in case of genuine entanglement $G$. For example, the maximal value of $G$ ($\sim$ 0.35) can be obtained for probability $P(\nu_{\mu} \rightarrow \nu_{e}) \sim  0.08$ in the SM. The LMA-Light scenario reduces the maximum value of $G$ to 0.3 for $P(\nu_{\mu} \rightarrow \nu_{e})\approx 0.06$. In case of IO, $G^{max} \approx 0.32$ can be achieved for $P(\nu_{\mu} \rightarrow \nu_{e}) \approx 0.065$ for SM interaction. On the other hand, for LMA-Dark sector $G$ as well as $P(\nu_{\mu} \rightarrow \nu_{e})$ are enhanced largely, such as, $G^{max} \approx 0.5$ for $P(\nu_{\mu} \rightarrow \nu_{e})\approx 0.115$. Also, the maximum value of flavor entropy $S$ remains in the range $2 - 2.5$ when $P(\nu_{\mu} \rightarrow \nu_{e})$ varies within $0.08 - 0.12$ for different scenarios.

\section{Conclusions}\label{Conclusions}
We study the effects of NSI on temporal correlations quantified in terms of LGtI as well as spatial correlations quantified in terms of flavor entropy and genuine tripartite entanglement in the oscillating neutrino system and compare the results with the SM scenario. We find that, in case of normal mass ordering, LGtI  violation of its classical bound in the presence of SM interaction is large in comparison to NSI. Conversely, in case of inverted mass ordering, LGtI-violation is enhanced for the  LMA-D scenario over the SM interaction. Similar  features have been observed in case of genuine entanglement measure, while the flavor entropy, which is a measure of residual entanglement, is not affected significantly. An interesting result of this work is that if LGtI is violated, $i.e., K_3$ exceeds its classical bound, at $E \approx 3$ GeV (energy corresponding to the maximum neutrino flux at DUNE), then this would be possible only for LMA-Dark solution with IO, thus pointing towards the existence of new physics. In the wake of observational implications, we have also presented correlation plots between oscillation probability and various correlation measures.

\subsection*{Acknowledgment}
The authors would like to thank Ashutosh Kumar Alok and Javid Naikoo for fruitful discussions.

\end{document}